\documentclass[prb,twocolumn,superscriptaddress,showpacs,%
preprintnumbers,amsmath,amssymb]{revtex4}
\usepackage{amsmath}
\usepackage{graphicx}
\usepackage{dcolumn}

\DeclareMathOperator{\sech}{sech}

\begin{document}

\title{Electrostatics of electron-hole interactions in van der Waals heterostructures}

\author{L. S. R. Cavalcante} 
\author{A. Chaves} \email{andrey@fisica.ufc.br}
\affiliation{Departamento de F\'isica, Universidade Federal do
Cear\'a, Caixa Postal 6030, Campus do Pici, 60455-900 Fortaleza,
Cear\'a, Brazil}
\author{B. Van Duppen}
\affiliation{Department of Physics, University of
Antwerp, Groenenborgerlaan 171, B-2020 Antwerp,
Belgium}
%\author{M. Z. Mayers}
%\affiliation{Department of Chemistry, Columbia University, New York, NY 10027, USA}
\author{F. M. Peeters}
\affiliation{Departamento de F\'isica, Universidade Federal do
Cear\'a, Caixa Postal 6030, Campus do Pici, 60455-900 Fortaleza,
Cear\'a, Brazil}
\affiliation{Department of Physics, University of
Antwerp, Groenenborgerlaan 171, B-2020 Antwerp,
Belgium}
\author{D. R. Reichman}
\affiliation{Department of Chemistry, Columbia University, New York, NY 10027, USA}

\date{ \today }

\begin{abstract}
The role of dielectric screening of electron-hole interaction in van der Waals heterostructures is theoretically investigated. A comparison between models available in the literature for describing these interactions is made and the limitations of these approaches are discussed. A simple numerical solution of Poisson's equation for a stack of dielectric slabs based on a transfer matrix method is developed, enabling the calculation of the electron-hole interaction potential at very low computational cost and with reasonable accuracy. Using different potential models, direct and indirect exciton binding energies in these systems are calculated within Wannier-Mott theory, and a comparison of theoretical results with recent experiments on excitons in two-dimensional materials is discussed. 
\end{abstract}

\pacs{}

\maketitle

\section{Introduction}

The physics of excitons and other electron-hole complexes in atomically thin materials \cite{Bhimanapathi, Tan, Basov, Mak, TonyReview, GlasovReview} has attracted great attention in the past few years, in part due to the high electron-hole binding energies observed in these systems, which are approximately ten times higher than those of conventional semiconductors, such as III-V and II-VI compounds, even when the latter are structured in quantum dots, wires or wells. \cite{TonyReview, Ugeda, Thygesen1, Tim} Excitonic Rydberg spectra of WS$_2$ \cite{Alexey} and WSe$_2$ \cite{He} have been measured in recent absorption experiments, where one can verify up to 3 excited states. These series, however, differ from that expected for a hydrogen-like electron-hole pair. Two-photon absorption measurements have also been used to investigate excitonic states with $p$-symmetry, where a slight degeneracy break with respect to $s$-states is expected.\cite{He} These features suggest that the electron-hole interaction potential in this system is not Coulombic: indeed, due to the lack of screening by the environment above the material layer, the interaction is expected to acquire a different form, as discussed decades ago  \cite{Rytova, Keldysh} in the context of thin semiconductor films.

The effective electron-hole interaction potential is straightforwardly found by analytically solving the Poisson equation for a dielectric slab surrounded by two media with different dielectric constants. This approach clearly provides a fully classical electrostatic description of the problem. It is far from guaranteed, however, that such a classical approach provides reasonable results in the limit of atomically thin materials, where quantum and dynamical effects may be sizeable. Using a classical effective potential to calculate exciton eigenenergies leads to a reasonable agreement between theory and experiment,\cite{Alexey} but only if additional screening due to the SiO$_2$ substrate in the experiment is taken into account. A more recently developed approach, \cite{Thygesen, Thygesen2, Thygesen3} involving quantum mechanical effects via ab initio calculations, is expected to provide better agreement in few layer cases, which has been confirmed by comparison to the same experimental results of Ref. \onlinecite{Alexey}. In this approach, known as the quantum electrostatic heterostructure (QEH) model, as well as the simple classical effective potential approach, the main effects of the environment on the electron-hole interactions are all included in the form of a static ($\omega = 0$) dielectric function. Dielectric functions for both approaches match for low wave vectors, but strongly disagree as $k$ increases, thus suggesting the QEH model captures important contributions to the dielectric function which are not captured by the simple classical effective model.

It is important to point out that despite the limitations of classical effective potential approaches \cite{Rytova, Keldysh} for describing atomically thin materials, they are a physical and efficient way of obtaining the electron-hole potential in the limit of a large number of layers. It is thus worthwile to investigate how this approach compares to the QEH model as the number of layers increases, in order to obtain a deeper understanding of the limitations of this simple approach. In the same spirit, it is important to compare both approximations for the case where substrate screening is important, as well as in the presence of layers of different materials, i.e. in van der Waals heterostructures. \cite{vdWreview}

In this paper, we explore the effective electron-hole interaction potential, suitable for charged particles in a $N$-layer vdW stack. This is accomplished by solving the Poisson equation for the potential experienced by a charged particle in a given layer due to a test charge placed in the same or any other layer. We demonstrate that such a classical electrostatic approach provides a very fast and computationally efficient means of achieving results which are surprisingly accurate when compared to those obtained from more sophisticated and expensive approaches based on ab initio calculations. Our results for the binding energy of inter-layer excitons in hetero-bilayers, as well as for intra-layer excitons in the presence of additional graphene capping layers, \cite{Archana} are discussed in light of recently reported experimental PL and absorption data for these systems. In addition, a detailed comparison is made with the recently developed QEH approach. \cite{Thygesen}

\section{Theoretical framework}

Theoretical approaches available in the literature for investigating electron-hole interactions in low dimensional systems surrounded by different dielectric media are usually based either on (i) classical electrostatics, where the interaction potential is obtained, e.g., by solving the Poisson equation for a stack of dielectric slabs, \cite{Rytova, Keldysh} or (ii) via direct or parametrized first principles calculations, the latter of which forms the basis of the recently proposed quantum electrostatic heterostructure approach, \cite{Thygesen} where the effective dielectric function of the vdW stack is obtained with the aid of ab initio-obtained density response functions of the separated layers that compose the heterostructure. In what follows, these two approaches are discussed in greater detail.  

\subsection{Electrostatic Transfer Matrix Method}

\begin{figure}[!b]
\centerline{\includegraphics[width=\linewidth]{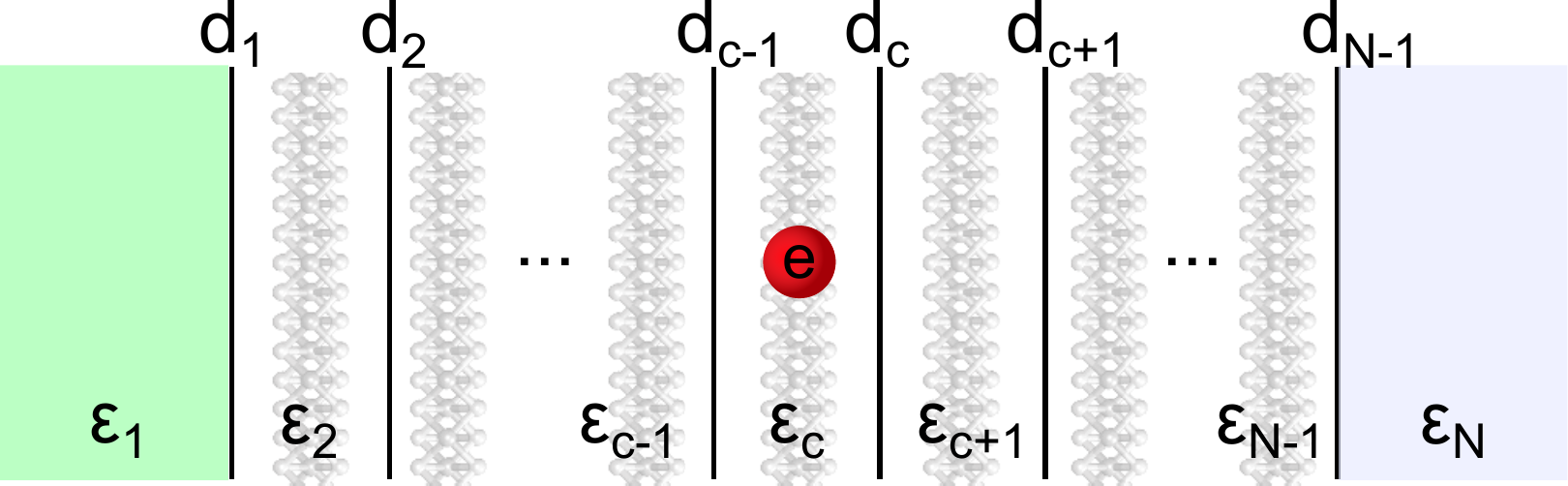}}
\caption{(Color online) Sketch of the series of interfaces between slabs with dielectric constants $\varepsilon_n$ describing each material layer. The charge ($e$) placed at the $c-$th slab generates a screened Coulomb potential at each layer, that obeys the Poisson equation with a space-dependent dielectric constant.} 
\label{fig:potentials}
\end{figure}

Let us assume a series of $N$ stacked layers along the $z$-direction, each with dielectric screening $\varepsilon_n$ ($n = 1, 2, ... N$), separated by interfaces at $z = d_n$ ($n = 1, ... N-1$), as sketched in Fig. 1. We take the origin as the center of the $c$-th layer, where the source charge is placed. Our aim is to calculate the potential at the $t$-th layer, where test charge is. For instance, spatially direct (indirect) excitons would have $c= t$ ($c \neq t$). For the $n$-th layer, the Poisson equation reads
\begin{equation}
\varepsilon^{\parallel}_n \nabla^2_{\rho,\theta}\Phi_{n,c} + \varepsilon^{\perp}_n\frac{\partial^2 \Phi_{n,c}}{\partial z^2} = q_n,
\end{equation}
where $q_n = -e\delta_{n,c}\delta(\vec r)$ is the point charge at this layer (which is non-zero only at the $c$-th layer). The negative sign implies we are assuming the source charge to be an electron.

The solution for the electrostatic potential at any layer $n$ is written in the form
\begin{eqnarray}
\Phi_{n,c}(\rho,z) = \frac{e}{4\pi\varepsilon_{c}\varepsilon_0}\int_{0}^{\infty}\left\{J_0(k\rho)\left[A_n(k)e^{kz} \nonumber \right.\right. \\
\left.\left. + B_n(k)e^{-kz} + e^{-k|z|}\delta_{n,c} \right]\right\}dk.
\end{eqnarray}
The electron-hole interaction potential $V^{t,c}_{eh} = e\Phi_{t,c}$ is more conveniently re-written as
\begin{equation}\label{eq.PhiFinal}
V^{t,c}_{eh}(\rho) = \frac{e^2}{4\pi\varepsilon_0}\int^{\infty}_0\frac{J_0(k\rho)}{\varepsilon_{t,c}(k)}dk,
\end{equation}
where the effective dielectric screening function for a hole in the $t$-th layer, at a distance $z_t$ from the point charge, is $\varepsilon_{t,c}(k) = \varepsilon_{c}\left[A_{t}(k)e^{kz_t} + B_{t}(k)e^{-kz_t} + \delta_{t,c}\right]^{-1}$. Notice that this expression covers both the direct ($z_t = 0$) and indirect exciton cases. We shall now look for a means of calculating $A_{t}(k)$ and $B_{t}(k)$.

Boundary conditions are imposed so that $B_1 \equiv 0$ and $A_{N} \equiv 0$, in order to avoid divergence as $z \rightarrow \pm \infty$. In addition, boundary conditions at each of the $N-1$ interfaces require continuity of the potential and its derivative, thus leading to a system of 2$(N-1)$ equations. Alternatively, one can represent each pair of equations for each interface in a matrix form
\begin{eqnarray}\label{eq.method}
{M}_n\left( \begin{array}{c} 
A_{n+1} \\ 
B_{n+1}
\end{array} \right)=\bar{M}_n\left( \begin{array}{c} 
A_{n} \\ 
B_{n}
\end{array} \right) - \left( \begin{array}{c} 
e^{kd_{c-1}} \\ 
\varepsilon_{c}e^{kd_{c-1}}
\end{array} \right)\delta_{n,c-1} \nonumber \\
 + \left( \begin{array}{c} 
e^{-kd_c} \\ 
-\varepsilon_{c}e^{-kd_c}
\end{array} \right)\delta_{n,c},
\end{eqnarray}
where
\begin{eqnarray}
\bar{M}_n = \left( \begin{array}{cc} 
e^{kd_n} & e^{-kd_n} \\ 
\varepsilon_{n}e^{kd_n} & -\varepsilon_{n}e^{-kd_n}
\end{array} \right), \nonumber \\
M_n\left( \begin{array}{cc} 
e^{kd_n} & e^{-kd_n} \\ 
\varepsilon_{n+1}e^{kd_n} & -\varepsilon_{n+1}e^{-kd_n}
\end{array} \right).
\end{eqnarray}

Combining all boundary conditions together yields
\begin{eqnarray}
\left( \begin{array}{c} 
0 \\ 
B_{N}
\end{array} \right)=\mathcal{M} \left( \begin{array}{c} 
A_1 \\ 
0
\end{array} \right) - \mathcal{M^\prime}\left( \begin{array}{c} 
e^{kd_{c-1}} \\ 
\varepsilon_{c}e^{kd_{c-1}}
\end{array} \right)\nonumber \\
 + \mathcal{M^{\prime\prime}}\left( \begin{array}{c} 
e^{-kd_c} \\ 
-\varepsilon_{c}e^{-kd_c}
\end{array} \right),
\end{eqnarray}
where $\mathcal{M} = {M}_{N-1}^{-1}\bar{M}_{N-1} \hdots {M}_{1}^{-1}\bar{M}_{1}$, $\mathcal{M}^{\prime} = {M}_{N-1}^{-1}\bar{M}_{N-1} \hdots {M}_{c}^{-1}\bar{M}_{c}{M}_{c-1}^{-1}$, and $\mathcal{M}^{\prime\prime} = {M}_{N-1}^{-1}\bar{M}_{N-1} \hdots {M}_{c+1}^{-1}\bar{M}_{c+1}{M}_c^{-1}$ can be seen as electrostatic transfer matrices (ETM). This allows us to solve for $A_1$ as
\begin{equation}
A_1 = \frac{(\mathcal{M}^{\prime}_{11} +\varepsilon_c\mathcal{M}^{\prime}_{12})e^{kd_{c-1}}-(\mathcal{M}^{\prime\prime}_{11}-\mathcal{M}^{\prime\prime}_{12}\varepsilon_{c})e^{-kd_{c}}}{\mathcal{M}_{11}}.
\end{equation}
Finally, once $A_1$ is obtained from the transfer matrices, $A_{t}(k)$ and $B_{t}(k)$ are calculated simply by applying the appropriate transfer matrices on $(A_1 \quad 0)^{T}$, according to Eq. (\ref{eq.method}).

\subsection{Quantum Electrostatic Heterostructure Model}

For the sake of completeness, here we briefly discuss the Quantum Electrostatic Heterostructure model for calculating the effective dielectric function in vdW stacks. More details concerning the derivation of this method are found in Ref. [\onlinecite{Thygesen}].

The QEH model uses in-plane averaged density response functions $\chi_i(k, \omega)$ that are obtained from ab initio calculations for each of the materials composing a van der Waals stack of layers. With a Dyson-like equation that couples the building blocks together via the Coulomb interaction, it is possible to calculate a full density response function $\chi_{ia, jb}$ that gives the magnitude of the monopole (dipole) density induced in the $i$th layer by a constant (linear) potential applied in the $j$th layer. Hence, the inverse dielectric matrix is obtained as

\begin{equation}
\epsilon_{ia, jb}^{-1}(k, \omega) = \delta_{ia, jb} + \sum_{lc}V_{ia, lc}(k)\chi_{lc, jb}(k, \omega),
\end{equation} 
where indices $i, j, l$ label the layers and $a, b, c = 0, 1$ correspond to monopole (0) and dipole (1) contributions. The Coulomb matrix is obtained from the potential $\Phi_{lc}(z, k)$ associated with the induced potential $\rho_{ia}(z, k)$, which is solution of a 1D Poisson equation, averaged over the thickness of the slab,

\begin{equation}
V_{ia, lc}(k) = \int \rho_{ia}(z, k) \Phi_{lc}(z, k) dz.
\end{equation}

Finally, an inverse Fourier transform of the potential,
\begin{equation}
V(k) = \sum_{ia, jb, lc}\rho^e_{ia}(k)\epsilon_{ia, jb}(k)^{-1}V_{jb, lc}(k)\rho^h_{lc}(k),
\end{equation}
results in the electron-hole potential in real space.

\subsection{Wannier-Mott Model}

Once the electron-hole potential is obtained from the methods described in the previous subsections, exciton eigenstates can be calculated within the Wannier-Mott model. \cite{Mostaani} The exciton Hamiltonian in this approach is given by
\begin{equation}\label{eq.Ham}
H = -\frac{1}{\mu_{ij}}\nabla_{2D}^2 - V^{t,c}_{eh}(\vec \rho),
\end{equation}
where $\mu_{ij} = \left(1/m_{e}^{i}+1/m_{h}^{j}\right)^{-1}$ is the reduced effective mass of the electron-hole pair, with an electron (hole) confined in the $i$-th ($j$-th) layer, $\vec \rho = \vec{\rho_e} - \vec{\rho_h}$ is the relative coordinate, and the center-of-mass contribution to the kinetic energy is taken to be zero. $V^{t,c}_{eh} (\vec{\rho})$ is the in-plane electron-hole interaction potential, calculated either by the QEH or the ETM methods. Energies and spatial coordinates are written in units of the Rydberg energy $R_y$ and the Bohr radius $a_0$, respectively. 

In the case of vdW heterostructures of TMDCs, which will be discussed in the following sections, the band offsets between the layers are finite, and thus the particles are able to tunnel between layers. Therefore, one should in principle consider, for each carrier, wave functions that are distributed across all layers, although with a much smaller probability in cases where band-offsets are large. The problem can then be treated as coupled quantum wells, described by a Hamiltonian matrix where the diagonal terms contain band offsets and in-plane potentials, whereas off-diagonal terms are hopping parameters. \cite{Jens, Zhang, Li} However, for the sake of simplification, we will assume the off-diagonal contributions to be small and the problem is then approximated by electrons and holes completely confined in individual layers. This approximation is reasonable, as demonstrated by the fact that recent DFT calculations \cite{Jens, Tony, Kaxiras} for vdW heterostructures show that their band structures at K (where the direct gap takes place and, consequently, the exciton is expected to be) is not significantly different from a superposition of the bands of their composing monolayer materials. This suggests that a quasi-particle Hamiltonian matrix for conduction and valence bands could be simply described each by a 2$\times$2 diagonal matrix, whose diagonal elements are just the monolayer bands, within a basis of completely layer-localized states. This situation supports the Hamiltonian in the form proposed in Eq. (\ref{eq.Ham}), which is then numerically diagonalized in order to provide the exciton binding energies shown in the following Sections.

\section{Results and discussion}

\subsection{Classical limits}

\begin{figure}[!b]
\centerline{\includegraphics[width=\linewidth]{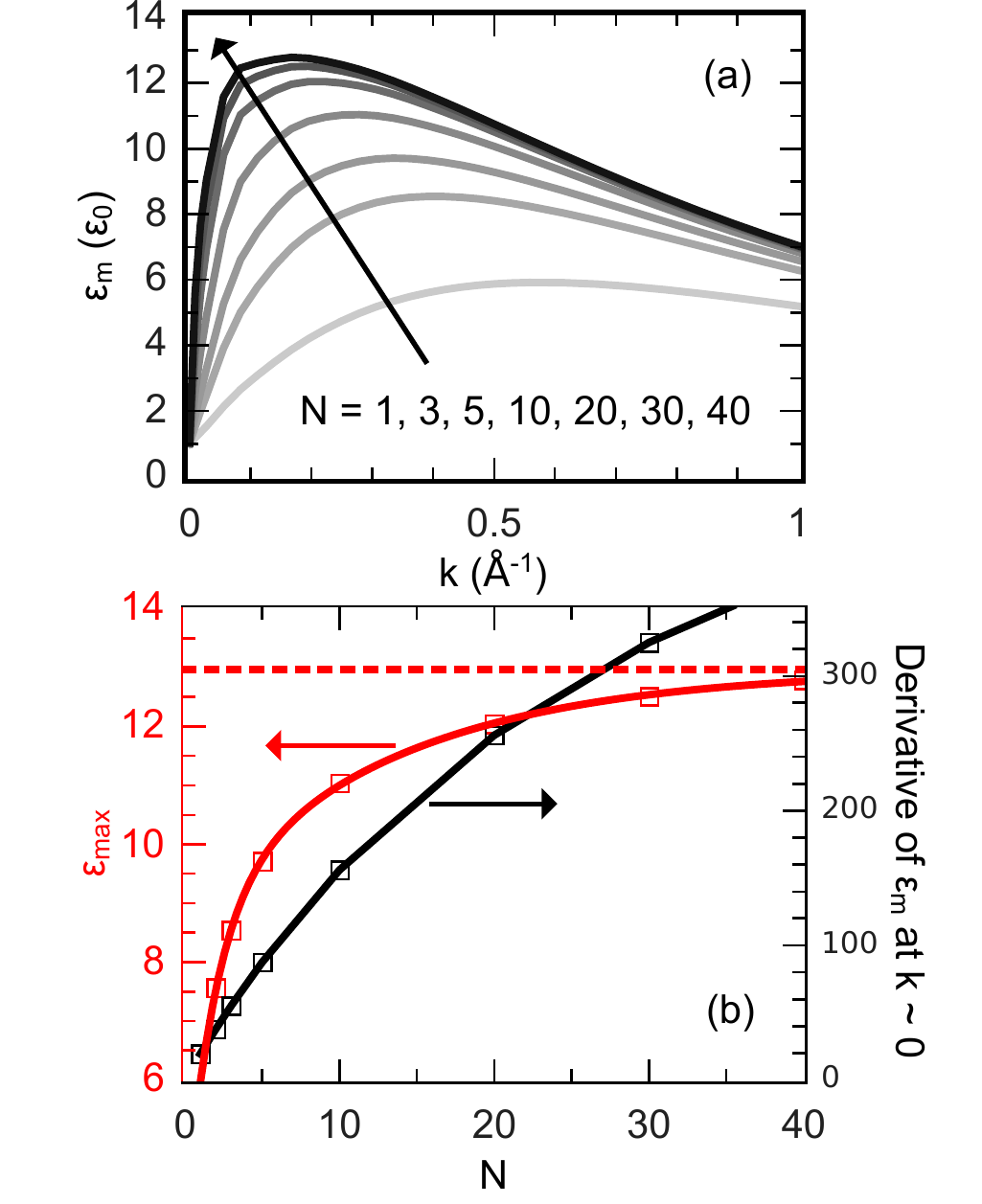}}
\caption{(Color online) (a) Average dielectric function for MoS$_2$ as calculated by QEH model for increasing number of layers. (b) Maximum value (red, left scale) of the curves shown in (a), along with their derivatives at $k = 0$ (black, right scale) as a function of the number of layers. Numerical results are shown as symbols. The curve on top of the $\epsilon_{max}$ (red) symbols is a fitting function (see text), whereas the one on top of the derivative results (black) is a guide to the eyes.} 
\label{fig:ThygesenLimits}
\end{figure}

Let us first investigate the limits of the effective dielectric functions of stacks of the same material, thus interpolating from the monolayer towards the bulk limit of a homogeneous system. An example is shown in Fig. \ref{fig:ThygesenLimits}(a), where results obtained by the QEH method for the macroscopic dielectric function \cite{Thygesen} of MoS$_2$ with $N = 1, 3, 5, 10, 20, 30,$ and 40 layers are illustrated. All curves exhibit a maximum $\varepsilon_{max}$, that increases with $N$ until it converges to a fixed value, as shown by (red) squares in Fig. \ref{fig:ThygesenLimits}(b), left scale. A fitting function for this maximum, $F(N) = A + Be^{-N/n_1} + Ce^{-N/n_2}$ is shown as a (red) solid curve, with $A = 12.96 \epsilon_0$, $B = -4.13\epsilon_0$, $C = -5.42 \epsilon_0$, $n_1 = 13.2$ and $n_2 = 1.9$. The most physically meaningful parameter in this case is $A$, which illustrates that for bulk MoS$_2$ (i.e. as $N \rightarrow \infty$), the dielectric function has a maximum value of approximately $\varepsilon \approx 12.96 \epsilon_0$. In addition, we expect that the low $k$ part of the dielectric function, which is an increasing function of $k$ for a finite number of layers, becomes negligibly small as the bulk limit is approached. In fact, the derivative of $\varepsilon_m$ at $k = 0$, shown as a function of $N$ as (black) squares (right scale) in Fig. \ref{fig:ThygesenLimits}(b), goes to infinity as $N \rightarrow \infty$. Both analyses suggest a dielectric function that converges to a dielectric constant $\varepsilon = 12.96 \epsilon_0$ as the bulk limit is reached, which agrees well with the dielectric constant of bulk MoS$_2$ found in the literature. \cite{Tim} The same procedure was done for other TMDCs, where we obtain the dielectric constants for bulk MoSe$_2$ ($\epsilon = 14.83 \epsilon_0$), WS$_2$ ($\epsilon = 11.74 \epsilon_0$), and WSe$_2$ ($\epsilon = 13.47 \epsilon_0$). This information will be used further in this Section for the ETM calculations of the electron-hole potential in vdW heterostructures.

As for the verification of the expected limits of the ETM method, let us use it to revisit the problem of a monolayer surrounded by two semi-infinite media, i.e. $N$ = 3. This problem was analytically solved by N. S. Rytova \cite{Rytova} and, later, by L. V. Keldysh, \cite{Keldysh} within some approximations, namely $\varepsilon_2 \gg \varepsilon_{1,3}$ and $d_2 - d_1 = d \ll a_0$. \cite{Keldysh} These approximations are such that for a charge in layer $c = 2$, the potential at layer $t = 2$ is given by 
\begin{eqnarray}
V^{R-K}_{eh} = \frac{e^2}{2\pi\varepsilon_0\varepsilon_2 d}\int^{\infty}_0\frac{J_0(k\rho)}{1 + \frac{\varepsilon_2 d}{\varepsilon_1 + \varepsilon_3}k}dk \nonumber\\
= \frac{e^2}{4\pi\varepsilon_0(\varepsilon_1+\varepsilon_3)\rho_0}\left[H_0\left(\frac{\rho}{\rho_0}\right) - Y_0\left(\frac{\rho}{\rho_0}\right) \right],
\end{eqnarray}
with $\rho_0 = \varepsilon_2d/(\varepsilon_1+\varepsilon_3)$, which is equivalent to Eq.(\ref{eq.PhiFinal}) with an effective dielectric function
\begin{equation}\label{eq.KeldyshEpsilon}
\varepsilon^{R-K}(k) = \frac{\varepsilon_1 + \varepsilon_3}{2}\left(1+\frac{d\varepsilon_2}{\varepsilon_1 + \varepsilon_3}k\right).
\end{equation}

In fact, for $N = 3$, after some algebra, our model yields
\begin{equation}\label{eq.EpsilonforN3}
\varepsilon(k) = \frac{{\varepsilon_1 + \varepsilon_3} + (1 + \frac{\varepsilon_1\varepsilon_3}{\varepsilon_2^2})\varepsilon_2\tanh(dk)
}{\left(1+\frac{\varepsilon_1\varepsilon_3}{\varepsilon_2^2}\right)+\left(1 - \frac{\varepsilon_1\varepsilon_3}{\varepsilon_2^2}\right)\sech(dk) + \frac{\varepsilon_1 + \varepsilon_3}{\varepsilon_2} \tanh(dk)}
\end{equation}
One can straightforwardly verify that Eq. (\ref{eq.KeldyshEpsilon}) is the $dk \rightarrow 0$, $\varepsilon_{1,3}/\varepsilon_{2} \rightarrow 0$ limit of Eq. (\ref{eq.EpsilonforN3}), as expected. Our model, thus, extends the classical approximation \cite{Rytova, Keldysh} to any value of dielectric constant and slab width, although the approximated linear dielectric function $\varepsilon^{R-K}(k)$ can still be seen as a low $k$ limit of $\varepsilon(k)$.

\begin{figure}[!t]
\centerline{\includegraphics[width=0.9\linewidth]{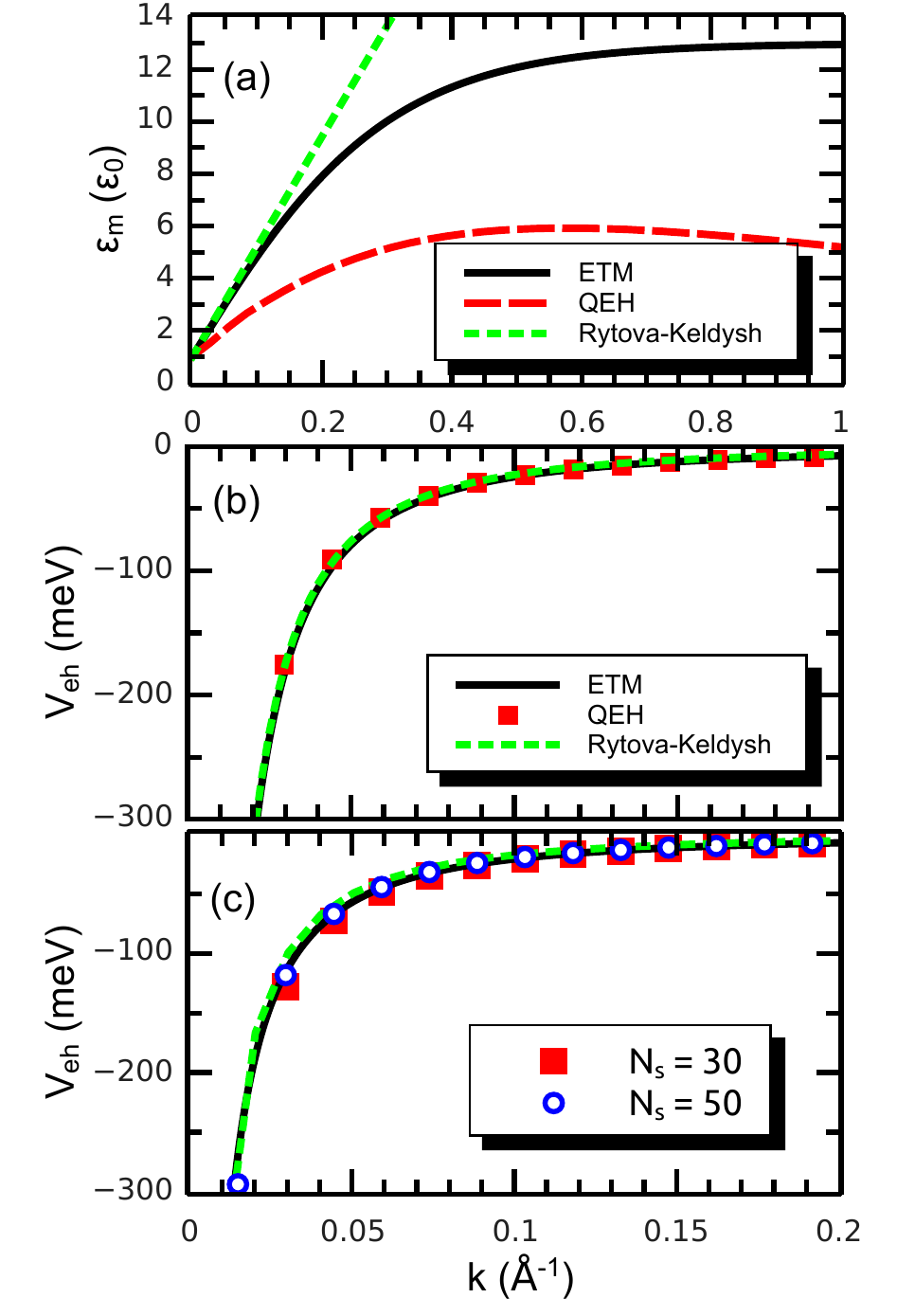}}
\caption{(Color online) (a) Effective dielectric function of a suspended monolayer MoS$_2$ as obtained by ETM and QEH methods, as well as with the Rytova-Keldysh effective potential approach. The effective interaction potential between electron and hole, as obtained by these methods, is shown in (b) and (c), for monolayer MoS$_2$ in the suspended case and over $N_s$ layers of BN substrate, respectively.} 
\label{fig:ETMLimits}
\end{figure}

The agreement between the effective dielectric functions of suspended monolayer MoS$_2$ obtained from the theory of Rytova and Keldysh and the ETM approach for low $k$ is verified in Fig. \ref{fig:ETMLimits}(a), which also shows the results obtained by the QEH method, illustrating somewhat worse agreement with these simpler approaches. Nevertheless, the effective interaction potential for both the suspended case (b) and for MoS$_2$ over a BN substrate (c), exhibit excellent agreement between all methods, including even the linear (Rytova-Keldysh) approximation for the dielectric function. For these calculations, we have assumed $\varepsilon_1 = 4 \epsilon_0$ (BN substrate), $\varepsilon_2 = 12.96 \epsilon_0$ (MoS$_2$) and $\varepsilon_3 = 1 \epsilon_0$ (vacuum), with $d_1 = -d_2 = 3.15$ \AA\,. Results for other TMDCs are qualitatively the same, and thus we will investigate only MoS$_2$ in what follows, unless otherwise explicitly stated. In addition, BN is chosen as the substrate (and in some cases capping) material because (i) of the similarity between its static dielectric constant and that of SiO$_2$, which has been commonly used as substrate in actual experiments, (ii) it is a layered material, which makes it suitable for the QEH calculations (although the ETM method allows for use of any kind of material, layered or not, as substrate or capping material), and (iii) because it has been used as capping material in some recent experiments. \cite{Li, Cui} Increasing the number of layers involved in the QEH calculations requires more computational memory, therefore one needs to limit the number of BN layers in the substrate. The QEH-obtained potential for MoS$_2$ over a BN substrate is shown as symbols in Fig. \ref{fig:ETMLimits}(c) for $N_s$ = 30 (red squares) and 50 (blue circles) BN layers. Indeed, increasing the number of BN layers renders the QEH-obtained potential closer to that of the ETM (black solid) one. 

\begin{figure}[!b]
\centerline{\includegraphics[width=0.9\linewidth]{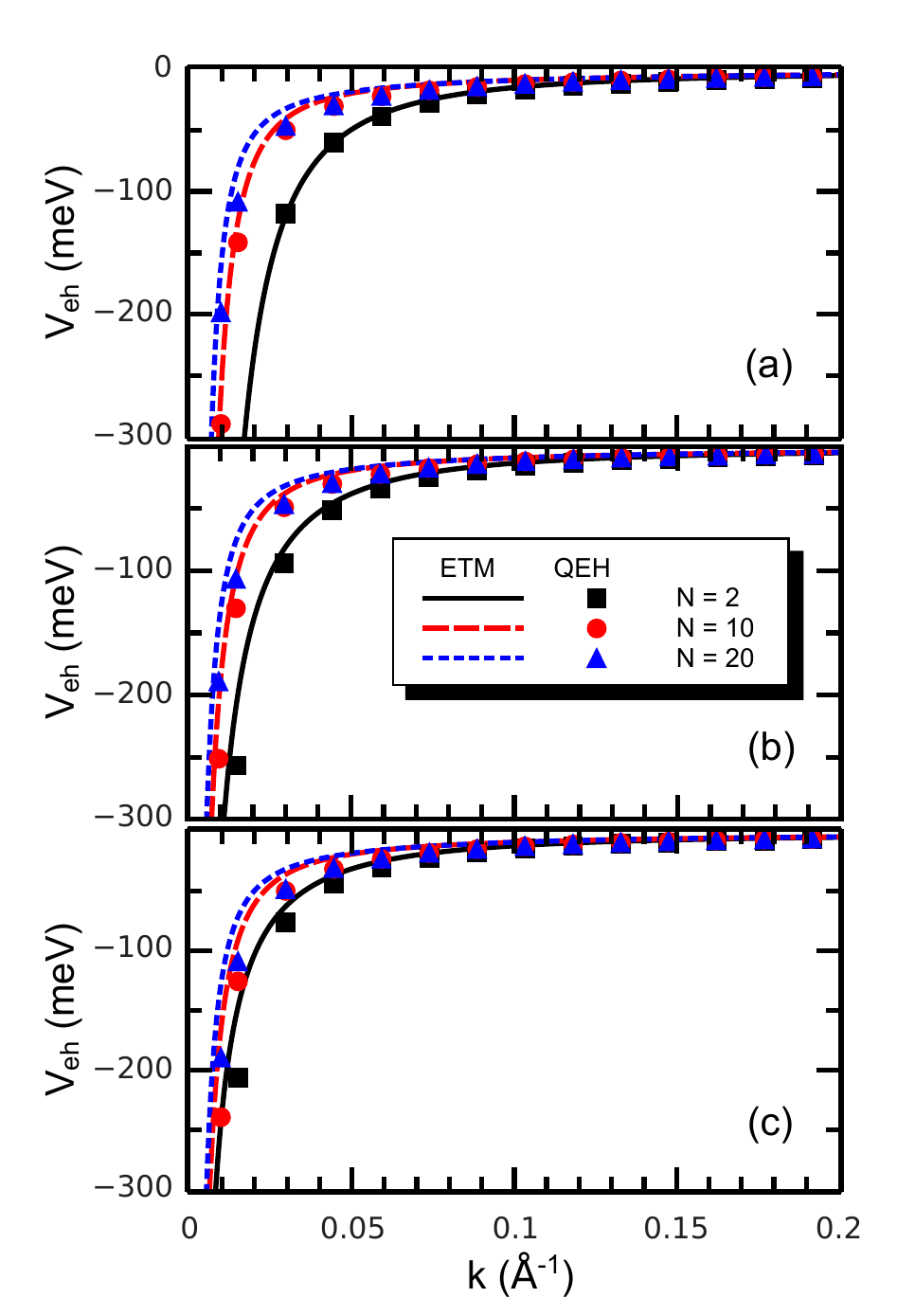}}
\caption{(Color online) Screened interaction potential between electron and hole, as obtained by QEH (symbols) and ETM (curves) methods, for N-layer MoS$_2$ (a) in the suspended case, (b) over a BN substrate, and (c) encapsulated by a BN substrate and a BN capping medium.} 
\label{fig:comparisonEncapsulated}
\end{figure}

The dependence of the screened electron-hole interaction potential on the number of MoS$_2$ layers is illustrated in Fig. \ref{fig:comparisonEncapsulated}, for (a) the suspended case, as well as for few layer MoS$_2$ (b) over a BN substrate, and (c) encapsulated by BN. In all cases, increasing the number of MoS$_2$ layers produces qualitatively the same effect in both QEH (symbols) and ETM (curves) methods. However, quantitative agreement between results from these two methods becomes somewhat worse as the number of layers increases. In the case of multi-layer MoS$_2$ over or encapsulated by BN, the lack of quantitative agreement is partially due to the small number of BN layers in the substrate and capping layers employed in our QEH calculations, which are taken as $N_s$ = 30 in the former case and $N_s$ = 15 (with 15 more BN capping layers) in the latter case. A larger number of BN layers, which would improve this agreement as previously discussed, is found to be very memory intensive when a large number of MoS$_2$ layers are considered, as in the $N = 20$ case.

\begin{figure}[!t]
\centerline{\includegraphics[width=0.9\linewidth]{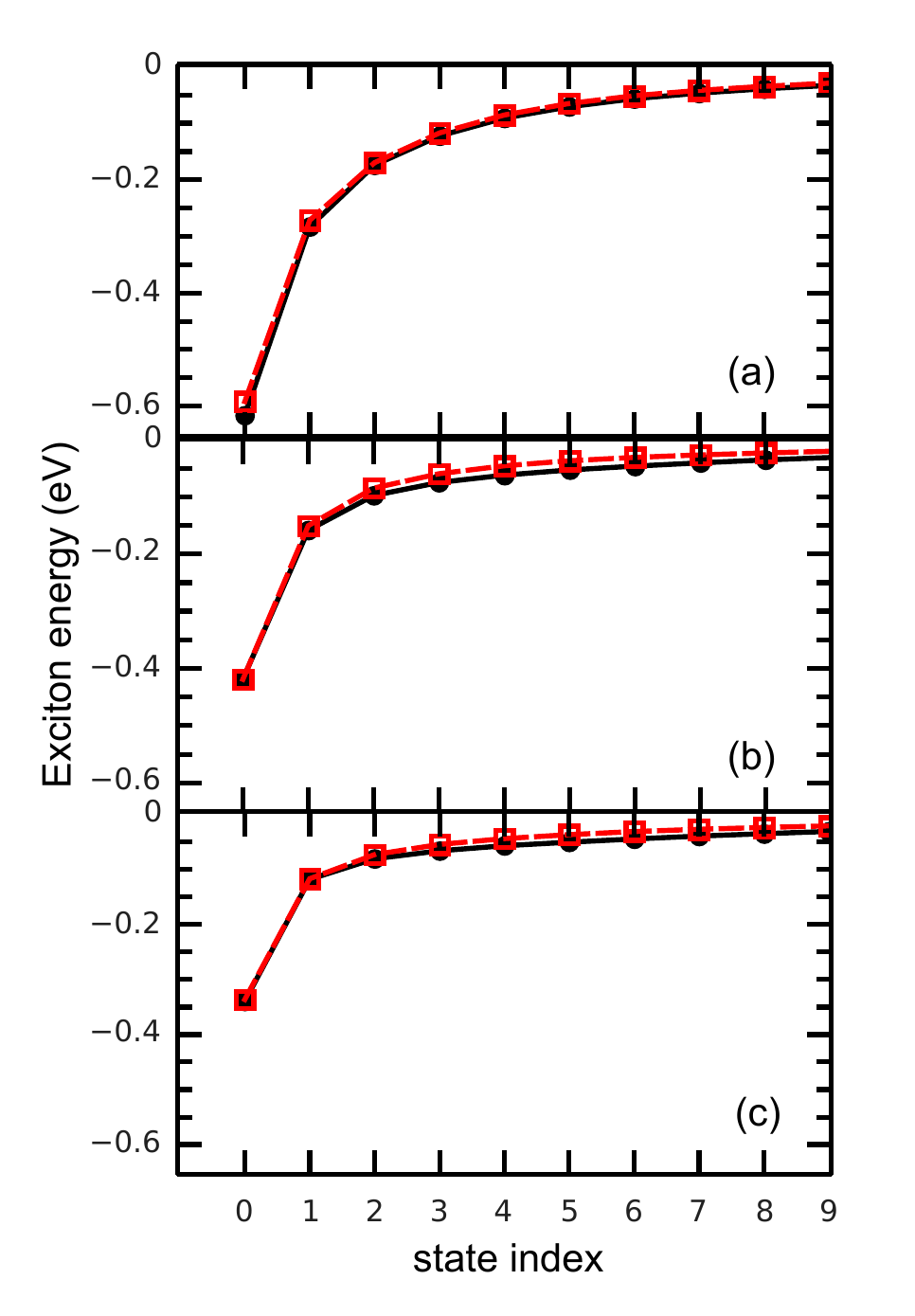}}
\caption{(Color online) Exciton energy states, as obtained by ETM (black full circles) and QEH (red open squares) methods, for monolayer MoS$_2$ (a) in the suspended case, (b) over a BN substrate, and (c) encapsulated by BN. Curves are guides to the eye.} 
\label{fig:energies1}
\end{figure}

The good agreement between these two methods for the monolayer case, especially for high values of $k$, suggests that low-lying exciton energy states, whose wave functions are narrower (wider) in real (reciprocal) space, as calculated by both approaches will also exhibit similar results. This is indeed verified in Fig. \ref{fig:energies1}, which shows the exciton state energies as obtained by ETM (black full circles) and QEH (red open squares) methods for (a) suspended monolayer MoS$_2$, as well as for this material (b) over a BN substrate and (c) encapsulated by BN. Ground state binding energies are found to be 0.616 eV in the suspended case, in good agreement with previous calculations, \cite{Thygesen, Tim}, whereas in the presence of a BN substrate, this energy is reduced to 0.419 eV and, when encapsulated by BN, it is further reduced to 0.336 eV, due to the additional dielectric screening by the surrounding environment. The difference between the two methods is almost zero for the ground state, but it reaches $\approx$ 13$\%$ for the 8th excited state of suspended MoS$_2$. Nevertheless, for all cases studied here, the highest energy difference found was $\approx$ 0.01 eV for highly excited states, which is within the accuracy limitations of usual experimental measurements of these states. 

\begin{figure}[!t]
\centerline{\includegraphics[width=0.9\linewidth]{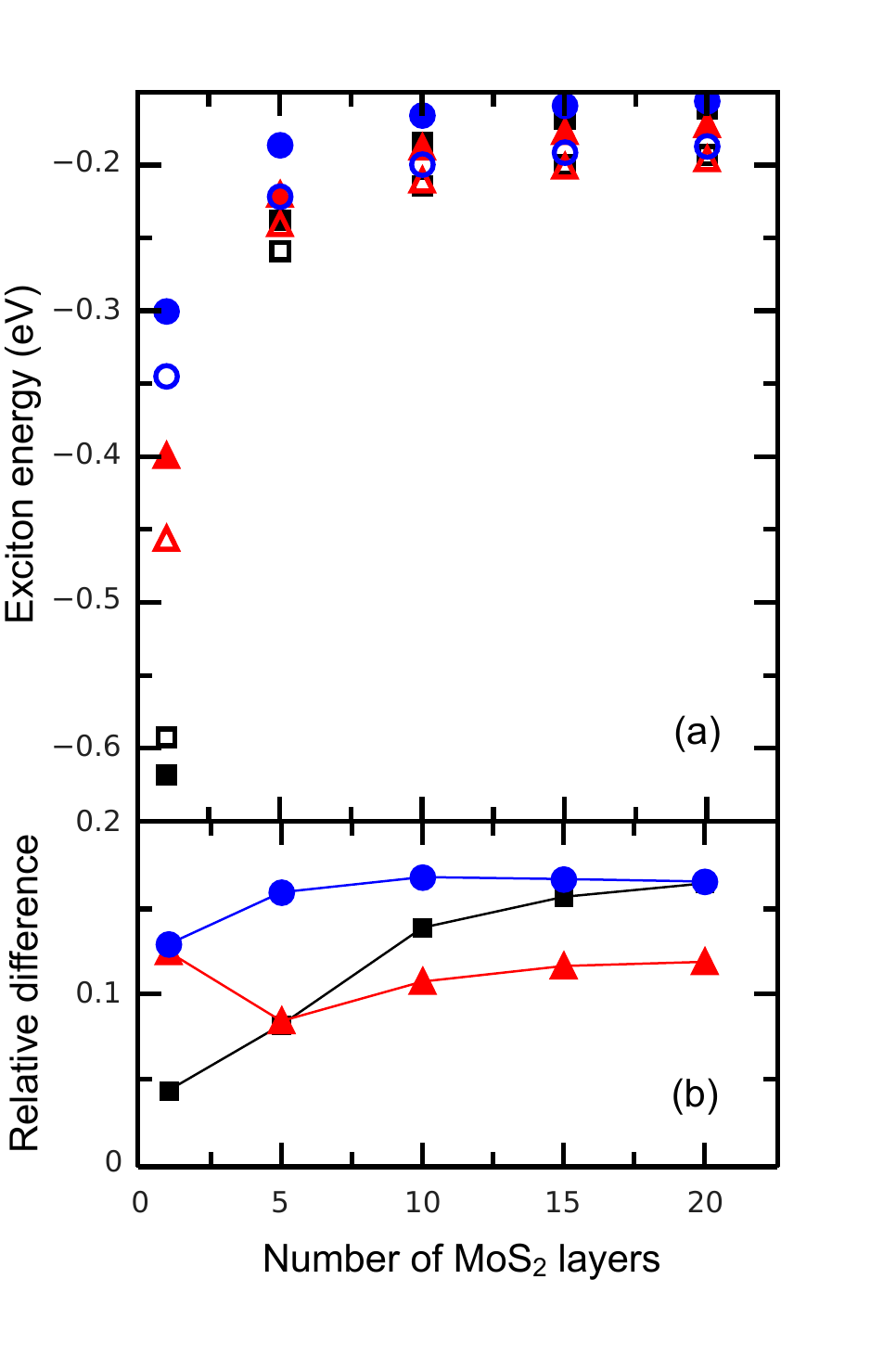}}
\caption{(Color online) (a) Exciton ground state energy as a function of the number of MoS$_2$ layers, as obtained by ETM (full symbols) and QEH (open symbols) methods, in the suspended case (black squares), over a BN substrate (red triangles), and encapsulated by BN (blue circles). (b) Relative difference between results obtained with the ETM and QEH methods. Lines are guides to the eye.} 
\label{fig:energiesN}
\end{figure}

In order to investigate the practical consequences of the observed difference between curves obtained with the ETM and QEH methods in the $N > 1$ case (see Fig. \ref{fig:comparisonEncapsulated}), we calculate the binding energy of a bound state composed of a positive and a negative charge in $N$-layer MoS$_2$. As this material acquires an indirect gap for $N \geq 2$, such a state is not relevant for excitonics, although it can still be used as a measure of the strength of the effective screened Coulomb interaction in the system which is relevant, e.g., for a charge-impurity bound state. Ground state binding energies are shown in Fig. \ref{fig:energiesN}(a) as a function of the number of MoS$_2$ layers in the suspended case (black squares), as well as for layers deposited on (red triangles) or encapsulated by BN (blue circles), as obtained by ETM (full symbols) and QEH (open symbols). Differences between methods (relative to the QEH results) are shown in Fig. \ref{fig:energiesN}(b) to be restricted to a range between 5$\%$ and 17$\%$. We point out that as the number of layers increases towards the bulk limit, the ETM method leads to the correct interaction potential. 

We now investigate how the ETM approach performs for an electron-hole interaction potential in two cases recently experimentally investigated, namely, a hetero-bilayer, i.e. a bilayer composed by two different TMDCs,\cite{vdWreview, Rivera, Rigosi, Fang, Chiu, Nayak, Zhu, Wurtzbauer, Calman} and a TMDC monolayer with extra dielectric screening due to a graphene capping layer. \cite{Archana} 

\subsection{Inter layer excitons in hetero-bilayers}

We have applied the theoretical model described in Sec. II to calculate exciton binding energies in vdW heterostructures consisting of the most common combinations of TMDCs experimentally investigated to date. Since a major focus in these systems is the study of inter-layer excitons, here we consider only heterostructures that exhibit a type-II band alignment, where this kind of exciton is energetically favorable. As part of the search for Bose-Einstein condensation of spatially polarized (inter-layer) excitons, recent studies \cite{Kezerashvili1, Kezerashvili2, Drummond} have investigated the binding energy of excitonic complexes in TMDCs double layers. In order to provide control of the inter-layer separation, the use of a few-layer BN spacer between the TMDCs that compose the vdW heterostructure has been proposed. \cite{vdWreview} 

\begin{figure}[!t]
\centerline{\includegraphics[width=\linewidth]{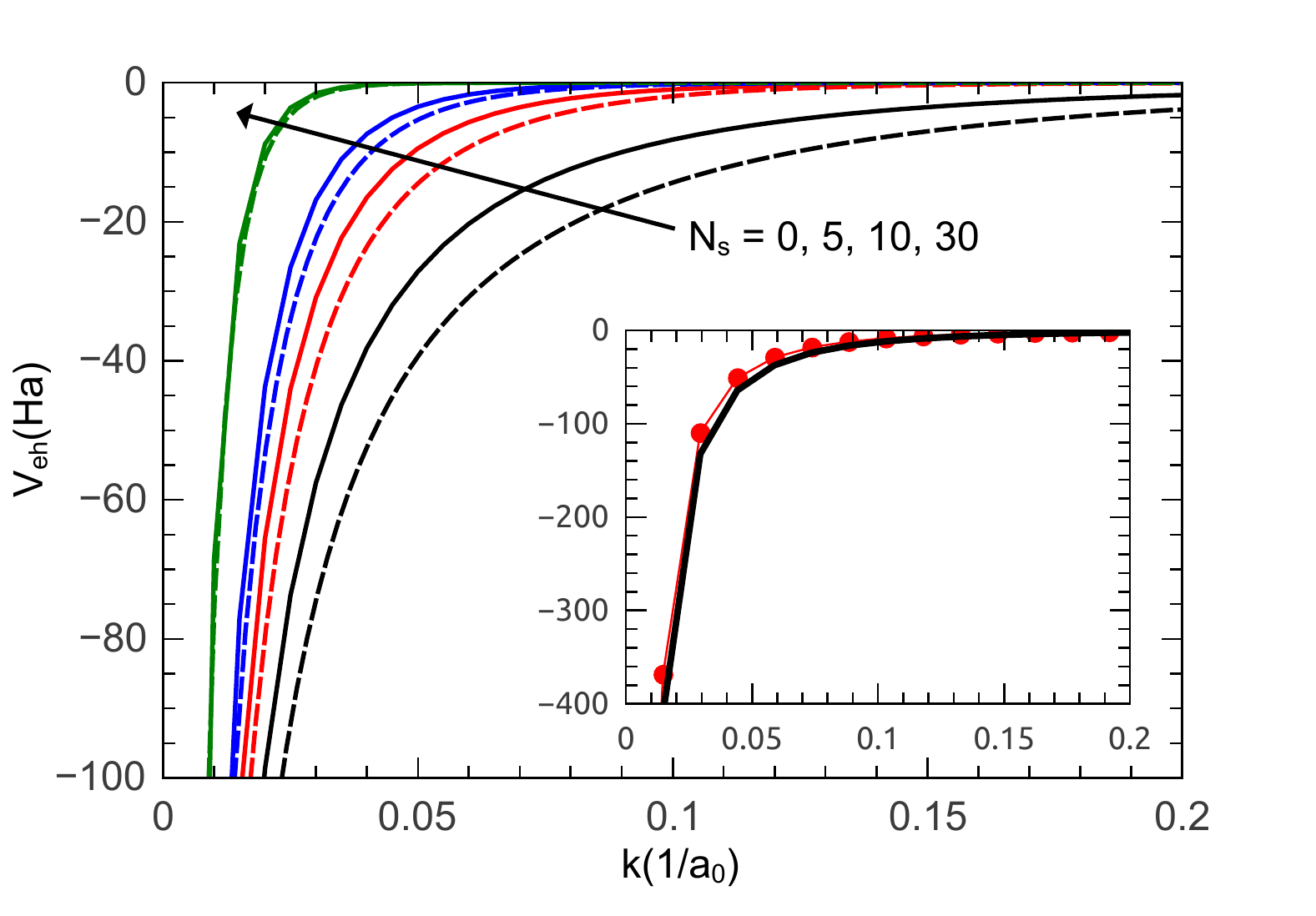}}
\caption{(Color online) Comparison between inter-layer electron-hole interaction potentials obtained by the ETM (solid) and the standard Coulomb form (dashed), for a MoS$_2$/WS$_2$ heterostructure encapsulated by BN and with a $N_s$-layer BN spacer between the TMDCs, for $N_s = 0$ (black), 5 (red), 10 (blue) and 30 (green). Inset: comparison between inter-layer electron-hole interaction potentials for suspended MoS$_2$/WS$_2$ hetero-bilayer, as obtained by ETM (black solid) and QEH (red symbols) methods.} 
\label{fig:interlayer}
\end{figure}

Previous calculations of excitonic complexes in these systems were mostly made under the approximation of a pure Coulomb interaction between electrons in one layer and holes in the other. The interaction potential in this case is given by $V_{Coulomb}(\rho) = -1\big/\epsilon_s \sqrt{\rho^2 + d_{z}^2}$, where $d_z$ is the distance between the center of the TMDC layers (where the charges are confined) and $\epsilon_s$ is the effective dielectric constant of the surrounding environment. In Fourier space, this potential is given by the expression $V_{Coulomb}(k) = -2\pi e^{-kd_z}/\epsilon_s k$. A comparison between this approximation and the actual potential obtained from solution of the Poisson equation by the ETM method for this combination of dielectric slabs is shown in Fig. \ref{fig:interlayer}. We consider a MoS$_2$/WS$_2$ heterostructure with a BN substrate, a BN capping medium, and a $N_s$-layer BN spacer between the TMDCs ($\epsilon_r$ = 4.4 $\epsilon_0$), to provide control of the distance between them. We point out that this encapsulation with BN is not necessary for actual heterostructures, but we consider it to enable the comparison with the same situation described by the recent use of the Coulomb approximation, where the possible difference between the dielectric constants of the inter-layer spacer, substrate and capping media has not been taken into account. We observe that interaction potentials obtained from the ETM (solid curves) assuming no BN spacer (i.e. $N_s$ = 0, black curves) are not well described by the Coulomb approximation (dashed curves). As the number of layers in the spacer increases to 5 (red), 10 (blue) and 30 (green), the results from these two approaches become more similar. This is reasonable, as the TMDCs layer thickness becomes less significant as compared to the BN media surrounding these layers. 

The ETM method and Coulomb approximation are both classical approaches for the inter-layer exciton problem. It is then important to compare the ETM results with the more sophisticated, ab inito based QEH method. Notice that calculations assuming BN as a surrounding environment and spacer would require a very large number of layers in QEH, which makes these calculations computationally expensive. We therefore investigate only the sample case of a suspended MoS$_2$/WS$_2$ hetero-bilayer with no BN spacer. Results for this case are shown in the inset of Fig. \ref{fig:interlayer}, where the ETM (QEH) obtained potential is shown as a black solid curve (red symbols). Potentials from both methods agree very well, and this is true for all combinations of TMDCs we investigated. As a measure of the consequences of the small difference between methods, we compare the exciton binding energies for MoS$_2$/WS$_2$, MoS$_2$/WSe$_2$, MoSe$_2$/WS$_2$ and MoSe$_2$/WSe$_2$. We obtain from the QEH method $E^{MoS_2/WS_2}_b = 281$ meV, $E^{MoS_2/WSe_2}_b = 271$ meV, $E^{MoSe_2/WS_2}_b = 279$ meV, and $E^{MoSe_2/WS_2}_b = 264$ meV, while ETM results overestimate these values by only 4$\%$, 8$\%$, 8$\%$ and 7$\%$, respectively. For the sake of simplicity, reduced effective masses are kept as $0.27m_0$ for all combinations, but numerical results will differ only by a few meV if the true values are considered. One conclusion is immediately drawn from these results: the inter-layer exciton binding energy for all combinations of TMDCs is of the order of $\approx 250$ - 300 meV, which is consistent with previous reports in the literature. \cite{Wilson, Tony, Thygesen2} This is important for the interpretation of experimentally observed photoluminescence peaks for vdW heterostructures. In order to substantiate that a given spectral peak observed in these experiments arises from such fully polarized inter-layer excitons, the energy of this peak needs to be consistent with the inter-layer quasi-particle gap, deduced by a binding energy of the appropriate order of magnitude. Nevertheless, we emphasize that our calculations were done assuming full electron-hole polarization, i.e. with each charge carrier confined exactly at a single layer, with no wave function projection on the other layer. This is expected to be the case for K-to-K point transitions in TMDCs hetero-bilayers. Recent experiments, \cite{Jens} however, suggest the presence of indirect (in reciprocal space) excitons associated with K-to-$\Gamma$ transitions, where holes are distributed across both layers, which naturally significantly increases the binding energy of these inter-layer excitons. 

Since the ETM provides a realistic inter-layer exciton potential at a low computational cost, it would be interesting to use this improved potential to revisit the problem of inter-layer excitons, trions and biexcitons discussed in the literature.\cite{Kezerashvili1, Kezerashvili2, Drummond} This is, however, outside of the scope of this paper and is left as a goal for future work.

\subsection{Dielectric screening due to a graphene capping layer}

In a recent experiment, \cite{Archana} capping a WS$_2$ monolayer with multi-layer graphene has been proposed as a way to provide control of the optical gap in the TMDC by engineering of the dielectric screening of the Coulomb interaction. It has been shown that the extra screening due to the graphene capping layer reduces the exciton binding energy, which is verified by the reduction of the energy difference between 1s and 2s states, observed as peaks in the reflectance spectrum around the A-exciton energy range. Although the optical gap of WS$_2$ is \textit{red}shifted after it is covered with graphene, we point out that the optical gap is comprised of a combination of this binding energy with the quasi-particle gap, which is also renormalized (reduced) via the change in the dielectric environment due to this graphene deposition. The separation between 1s and 2s peaks, however, is unaffected by the quasi-particle gap renormalization, therefore, its reduction after deposition of graphene is a measure of the enhanced dielectric screening of the electron-hole interactions in the WS$_2$ exciton state.

\begin{figure}[!b]
\centerline{\includegraphics[width=\linewidth]{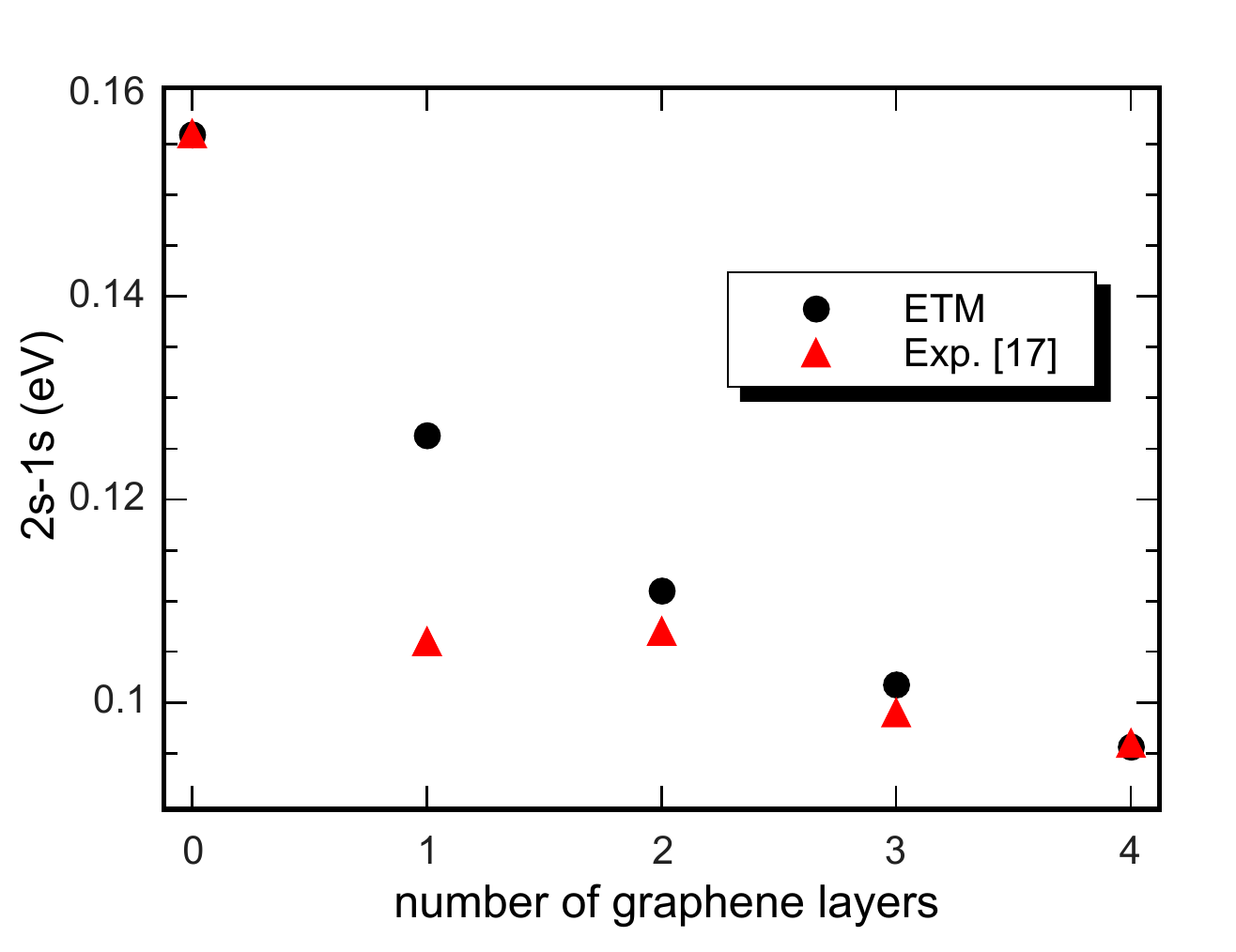}}
\caption{(Color online) Difference between ground (1s) and first excited (2s) exciton states in WS$_2$ as calculated by ETM (black circles). Experimental values for this system \cite{Archana} are shown as red triangles.} 
\label{fig:graphene}
\end{figure}

Figure \ref{fig:graphene} shows the 1s-2s separation for exciton states of monolayer WS$_2$, as a function of the number of graphene layers deposited. In order to obtain the correct 1s-2s separation for bare WS$_2$ as compared to the experiment, we had to assume a substrate with dielectric constant around $7 \epsilon_0$, which is higher than that of SiO$_2$, the actual substrate in the sample \cite{Archana}. We assume each deposited graphene layer to have a 3.36 \AA\, thickness and the same dielectric constant as graphite ($\epsilon_g = 10 \epsilon_0$), as required by the ETM method. Numerical results (black circles) agree very well with the experimental data (red triangles), thus validating the ETM method as a powerful tool to investigate the tuning of exciton peaks in Coulomb-engineered systems.

\section{Conclusions}

We have proposed a classical (electrostatic) model for describing the electron-hole interaction potential in few layer TMDCs and their vdW heterostructures. With its transfer matrix-like structure, the method developed here is easily manipulated to calculate the screened electron-hole interaction potential in any combination of TMDCs layers and substrates for either spatially direct (intra-layer) or indirect (inter-layer) excitons. We verify this method correctly converges to the standard effective potential of Rytova and Keldysh in the limit of small thickness and large differences between dielectric constants. It also yields the ordinary Coulomb potential for an inter-layer electron-hole interaction if the layers in which the charges are confined are separated by a large distance. A comparison between the proposed electrostatic transfer matrix method and the recently developed ab initio-based quantum electrostatic heterostructure (QEH) method \cite{Thygesen} is performed, where semi-quantitative agreement between results from both methods is demonstrated. Results from the ETM method are demonstrated to be very accurate for the exciton ground state and reasonably accurate (up to 0.01 eV error) for excited states, in comparison with those from the QEH method. Worse accuracy is observed in the case of inter-layer excitons in hetero-bilayers, where the difference in ground state binding energies may reach 0.02 eV ($\approx 8 \%$). Nevertheless, by paying the price of somewhat lower accuracy, the ETM method requires much lower computational overhead and an input based only on the dielectric constants of the bulk parent materials, in contrast to the input required by other DFT-based methods. By providing a facile and inexpensive means of obtaining the interaction potential, the ETM proves to be a powerful tool for calculations where interactions between charges need to be computed numerous times, such as in Monte Carlo based techniques for studying many-particle states. \cite{Drummond, Matt, Needs, Mostaani, vanderDonck}

Inter-layer exciton binding energies are found to be around $\approx 250$-300 meV, which is substantially lower than those of intra-layer excitons in monolayer TMDCs, $\approx$ 550 meV. \cite{Tim} This result is of importance in the interpretation of photoluminescence peaks in experiments involving vdW heterostructures. We have also successfully applied our method in the modelling of recently observed Coulomb engineered exciton states in WS$_2$ capped by few-layer graphene. \cite{Archana} 

We believe the fast and highly adjustable method developed here will be of use for verification, interpretation or prediction of excitonic peak positions in future experiments involving light-matter interactions in vdW stacks of layered materials. Work using the ETM approach to investigate excitons in inter-layer situations is currently under way.

\acknowledgments  Discussions with A. Chernikov and A. Raja are gratefully acknowledged. This work has been financially supported by CNPq, through the PRONEX/FUNCAP, PQ and Science Without Borders programs, and the FWO-CNPq bilateral program between Brazil and Flanders. BVD acknowledges support from the Flemish Science Foundation (FWO-Vl) by a postodoctoral fellowship. DRR was supported by NSF CHE-1464802.

\textit{Note added:} After this work was complete, we became aware of two related studies with some overlap with the discussions in the present paper: Ref. [\onlinecite{TimNew}] focus on the effect of the dielectric environment on the optical and electronic properties of monolayer MoS$_2$, whereas Ref. [\onlinecite{KochNew}] calculates bandgap renormalization of a TMDC from the monolayer to the bulk limits, combining the bulk DFT-obtained dielectric tensor and a massive Dirac fermion model.

\end{document}